# Hybrid skin-topological effect induced by eight-site cells and arbitrary adjustment of the localization of topological edge states

Jianzhi Chen(陈健智)[1], Aoqian Shi(史奥芊)[1], Yuchen Peng(彭宇宸)[1], Peng Peng(彭鹏)[1], and Jianjun Liu(刘建军)[1,2*]

[1]Key Laboratory for Micro/Nano Optoelectronic Devices of Ministry of Education, School of Physics and Electronics, Hunan University, Changsha 410082, China
[2]Greater Bay Area Institute for Innovation, Hunan University, Guangzhou 511300, China

*Corresponding author. Email: jianjun.liu@hnu.edu.cn

The hybrid skin-topological effect (HSTE) in non-Hermitian systems exhibits both the skin effect and topological protection, offering a novel mechanism for the localization of topological edge states (TESs) in electrons, circuits, and photons. However, it remains unclear whether the HSTE can be realized in quasicrystals, and the unique structure of quasicrystals with multi-site cells may provide novel localization phenomena for TESs induced by the HSTE. We propose an eight-site cell in two-dimensional quasicrystals and realize the HSTE with eight-site nonreciprocal intracell hoppings. Furthermore, we can arbitrarily adjust the eigenfield distributions of the TESs and discover domain walls associated with effective dissipation and their correlation with localization. We present a new scheme to precisely adjust the energy distribution in non-Hermitian quasicrystals with arbitrary polygonal outer boundaries.

Compared with Hermitian systems, non-Hermitian systems are more suitable for simulating open physical systems in real life; thus, extensive research has been conducted on various aspects of these systems, including non-Bloch bulk-boundary correspondence,[1–7] non-Bloch topological invariants,[1,3,4,6,8,9] generalized Brillouin zone,[1,4,8,10] non-Hermitian skin effect,[1,3,6–15] exceptional points (EPs),[1,4,5,13,16] and higher-order topology.[12,17,18] To describe the topological properties of both Hermitian and non-Hermitian systems in real space, different topological invariants, such as the Chern number,[8,9,19–21] Bott index,[9,22–24] and quantized quadrupole moment $Q_{xy}$,[25–27] have been analyzed. The study of non-Hermitian systems has been extended beyond electronic models to encompass other physical systems, such as cold atoms,[28] circuits,[11,29] optics,[30–37] acoustics,[38,39] and thermal diffusion,[13,15,40] providing theoretical bases[11,13,15,28,30–40] and experimental verifications[29,38,39] for the design of related devices. Recently, the first observation of the non-Hermitian skin effect in a photonic crystal showed the feasibility and potential of incorporating the topological edge state (TES) and skin effect in photonic systems.[33]

In recent years, researchers have realized the hybrid skin-topological effect (HSTE), a type of skin effect that can only be observed in TESs rather than bulk states, induced by nonreciprocity[28,29,41–43] or gain and loss[33–36,44–46] in electronic,[41–46] cold-atom,[28] circuit,[29] and optical systems.[33–36] Currently, studies on the HSTE are primarily focused on periodic systems[28,29,33–36,41–45] or hyperbolic lattices.[46] However, it remains unclear whether the HSTE can be realized in quasi-periodic systems, and the localization effects and physical laws of these systems are still not well understood. So far, quasicrystals have attracted attention because of their realization and features of topological states[22,23,25,26,47] and non-Hermitian skin effects.[48,49] In addition, only threefold, fourfold, and sixfold rotation symmetries exist in periodic crystals, which hinder us from controlling the energy distribution discretionarily. Quasicrystals can show some other symmetries that are forbidden in crystals. These symmetries may help to realize special phenomena and applications. Furthermore, a cell with *m* sites corresponding to two-dimensional quasicrystals with *n*-fold rotation symmetry (where *n/m* is an integer) offers great flexibility for the realization and utilization of the localization of topological states.[25,26,50,51] The HSTE in quasicrystals may act as a topological switch by designing integrated circuits to control the electrical signal propagations.[29,52]

In this Letter, we study the HSTE in two-dimensional quasicrystals, realized via eight-site nonreciprocal intracell hoppings, and propose two approaches to modify the localization of the TESs using HSTE in non-Hermitian systems, enabling arbitrary adjustment of the eigenfield distributions of the TESs. Furthermore, our findings reveal the presence of domain walls associated with the effective dissipation in the HSTE induced by nonreciprocal intracell hoppings. We present novel strategies for the construction of localized modes in non-Hermitian systems and for accurate control of the energy distribution in quasicrystals.

We utilize the Ammann–Beenker (AB) tiling quasicrystal with eightfold rotation symmetry and establish a regular octagonal outer boundary, as shown in Fig. 1(a). The AB tiling quasicrystal comprises numerous squares and rhombi with two 45° angles. Each point represents a cell in Fig. 1(a), where each cell contains eight sites, as shown in Fig. 1(b). In analogy to the two-dimensional Su–Schrieffer–Heeger (SSH) model with square lattices, in which the HSTE can be realized,[41] we introduce nonreciprocal intracell hoppings (the nonreciprocity is achieved using a non-zero $\delta$, with $\delta$ quantifying the nonreciprocal hopping between two sites) between the eight sites in the cells. In addition, we introduce intercell hoppings on the sides and diagonals of the square units, as well as on the sides and shorter diagonals of the rhomboid units, and the schematics of these intercell hoppings are shown in Figs. 1(c) and 1(d), respectively.

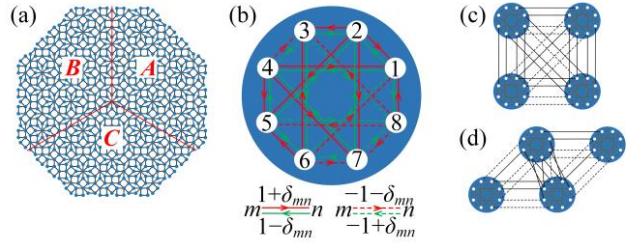

**Fig. 1** (a) AB tiling quasicrystal with a regular octagonal outer boundary. The inner boundary is set for calculation of the Chern numbers of the bulk bands with the Kitaev formula. (b) A schematic of the cell, and two schematics of the reciprocal intercell hoppings: (c) a square unit and (d) a rhomboid unit. The solid lines represent the positive hopping, while the dashed lines represent the negative hopping. Depending on the outer boundary that is selected, the rhomboid unit may be partially outside the boundary; in this case, the triangular part of the rhomboid unit that is outside the boundary should be removed.

The Hamiltonian *H* of the system is:
$$H = H_{\text{intra}} + H_{\text{inter}}, \tag{1}$$
where the intracell hopping is represented by $H_{\text{intra}}$, and the intercell hopping is represented by $H_{\text{inter}}$,
$$H_{\text{intra}} = \sum_j c_j^\dagger T c_j, \tag{2}$$
$$H_{\text{inter}} = \sum_{j \neq k} c_j^\dagger t f(r_{jk}) \Omega(\phi_{jk}) c_k, \tag{3}$$
where $c_j^\dagger = (c_{j1}^\dagger, c_{j2}^\dagger, ..., c_{j7}^\dagger, c_{j8}^\dagger)$ and $c_j = (c_{j1}, c_{j2}, ..., c_{j7}, c_{j8})$ are the electron creation operator and electron annihilation operator in the cell *j*, respectively, *T* is the intracell hopping in a cell, *t* is the intercell hopping amplitude, $f(r_{jk})$ is the spatial decay factor of the intercell hopping amplitude, and $\Omega(\phi_{jk})$ is a term related to the angle between cells. The detailed expressions of the physical quantities in Eqs. (2) and (3), including $\delta$ in Fig. 1(b), are provided in Section A of the Supplementary Material.

The topological properties of disordered structures can be studied using the Kitaev formula,[19−21,46] which is suitable for calculating the real-space Chern numbers:

$$C_\alpha = 12\pi i \sum_{j \in A} \sum_{k \in B} \sum_{l \in C} (P^\alpha_{jk} P^\alpha_{kl} P^\alpha_{lj} - P^\alpha_{jl} P^\alpha_{lk} P^\alpha_{kj}). \tag{4}$$

As shown in Fig. 1(a), the quasicrystal is divided into three neighboring regions arranged in counterclockwise order (regions $A$, $B$, and $C$), with sites $j$, $k$, and $l$ in regions $A$, $B$, and $C$, respectively. Here, $\alpha$ is the label of the bulk band, and $P^\alpha$ is the bulk-band projection operator for non-Hermitian systems:[8,9,19−21,46]

$$P^\alpha = \sum_{n \in \alpha} |nR\rangle \langle nL|, \tag{5}$$

and $P^\alpha_{jk}$ is an element of $P^\alpha$:

$$P^\alpha_{jk} = \sum_{n \in \alpha} \langle j | nR \rangle \langle nL | k \rangle, \tag{6}$$

where the right eigenstates $|nR\rangle$ satisfy $H|nR\rangle = E_n|nR\rangle$, and the left eigenstates $|nL\rangle$ satisfy $H^\dagger |nL\rangle = E_n^* |nL\rangle$; $|nR\rangle$ and $|nL\rangle$ are orthonormal. Irrespective of the arrangement of the adjacent regions $A$, $B$, and $C$, $C_\alpha$ remains a constant value.[19,20] The details of the calculation of the Chern numbers using the Kitaev formula, as well as an alternative calculation method (local Chern marker), are provided in Sections B and C of the Supplementary Material, respectively. TESs emerge in the bulk gaps when the Chern numbers are non-zero.

The non-Hermitic skin effect can be characterized by the winding number[1,3,6−12,14,15] or auxiliary Hamiltonian.[12,45] In a periodic system, the auxiliary Hamiltonian is

$$\tilde{H}(\bm{k}, E_r) = \begin{pmatrix} 0 & H(\bm{k}) - E_r \\ H^\dagger(\bm{k}) - E_r^* & 0 \end{pmatrix}, \tag{7}$$

where $E_r$ is the reference energy and corresponds to an eigenvalue of the TESs. Here, $\tilde{H}(\bm{k}, E_r)$ is Hermitian. The auxiliary Hamiltonian can be understood from the bulk-boundary correspondence in the model with chiral symmetry,[5] while the relationship between the winding number of the vector {Re[$H(\bm{k}) - E_r$], Im[$H(\bm{k}) - E_r$]} and the edge states of $\tilde{H}(\bm{k}, E_r)$ is similar to that in the SSH model.[53] For a one-dimensional system, the existence of the topological zero modes for the auxiliary Hamiltonian $\tilde{H}(\bm{k}, E_r)$ is protected by the winding of the vector {Re[$H(\bm{k}) - E_r$], Im[$H(\bm{k}) - E_r$]}. Hence, for a two-dimensional system, if the non-Hermitian Hamiltonian $H(\bm{k})$ is topologically nontrivial for $E_r$ and the skin effect occurs, second-order zero-energy corner states emerge under the open boundary conditions (OBC) for the auxiliary Hamiltonian $\tilde{H}(\bm{k}, E_r)$.[45] To study the non-Hermitian skin effect in two-dimensional quasicrystals, and by considering the reciprocal intercell hopping, Eq. (7) can be transformed into a real-space Hamiltonian:

$$\tilde{H}(E_r) = \tilde{H}_{\text{intra}}(E_r) + \tilde{H}_{\text{inter}}, \tag{8}$$

where

$$\tilde{H}_{\text{intra}}(E_r) = \sum_j c_j^\dagger \begin{pmatrix} 0 & T - E_r \\ T^\dagger - E_r^* & 0 \end{pmatrix} c_j, \tag{9}$$

$$\tilde{H}_{\text{inter}} = \sum_{j \neq k} c_j^\dagger \begin{pmatrix} 0 & tf(r_{jk})\Omega(\phi_{jk}) \\ tf(r_{jk})\Omega(\phi_{jk}) & 0 \end{pmatrix} c_k, \tag{10}$$

where $c_j^\dagger = (c_{j1}^\dagger, c_{j2}^\dagger, ..., c_{j15}^\dagger, c_{j16}^\dagger)$ and $c_j = (c_{j1}, c_{j2}, ..., c_{j15}, c_{j16})$; $\tilde{H}_{\text{intra}}(E_r)$ and $\tilde{H}_{\text{inter}}$ represent the intracell and intercell hopping of $\tilde{H}(E_r)$, respectively. By investigating the presence of zero-energy corner states for $\tilde{H}(E_r)$, we can ascertain the occurrence of the skin effect on the TESs in two-dimensional quasicrystals.

Although the model studied here is not a gain–loss model, we can still refer to the effective dissipation[54] associated with the chiral skin effect in gain–loss models to analyze the domain walls and determine the localization. The analysis of domain walls necessitates the definition of a global dissipation $\gamma_{\text{global}}$ under a periodic boundary condition (PBC):

$$\gamma_{\text{global}}(x) = \gamma(x) - \bar{\gamma}, \tag{11}$$

where $\gamma(x)$ is the dissipation, and $\bar{\gamma}$ is the average value of the inhomogeneous dissipation. If $x_0^+$ ($x_0^-$) satisfies $\gamma_{\text{global}}(x_0^+ + 0^-) < 0$ [$\gamma_{\text{global}}(x_0^- + 0^-) > 0$] and $\gamma_{\text{global}}(x_0^+ + 0^+) > 0$ [$\gamma_{\text{global}}(x_0^- + 0^+) < 0$] exists, then $x_0^+$ ($x_0^-$) has an A-type (B-type) global dissipative domain wall, and a chiral mode with negative (positive) velocity will be localized at $x_0^+$ ($x_0^-$).

The Hamiltonian $H$ with inhomogeneous dissipation can break down into:

$$H = H' + \delta H_{\text{D}}, \tag{12}$$

where $H'$ represents the part of the Hamiltonian unrelated to dissipation, and $\delta H_{\text{D}}$ represents the part of the Hamiltonian related to dissipation. In this study, the dissipation is induced by the nonreciprocal hoppings (nonzero $\delta$). The details of $\delta H_{\text{D}}$ are provided in Section A of the Supplementary Material.

Here, $\gamma_{\text{eff}}$ represents the effective dissipation of the TESs, defined as

$$\gamma_{\text{eff}} = \text{Im}\langle \psi_{\text{edge}} | \delta H_{\text{D}} | \psi_{\text{edge}} \rangle, \tag{13}$$

where $\psi_{\text{edge}}$ is the wave function of the TESs. By calculating the effective dissipation $\gamma_{\text{eff}}$ of each open boundary separately in the $x_{\text{PBC}}$–$y_{\text{OBC}}$ coordinate system (the $x$ and $y$ axes are relative to the PBC and OBC, respectively, i.e., the axis parallel to the periodic boundary is the $x$-axis, and the axis parallel to the open boundary is the $y$-axis), and comparing it with the corresponding average value of the inhomogeneous dissipation $\bar{\gamma}$ [as shown in Eq. (11)] to obtain the global dissipation $\gamma_{\text{global}}$ of each edge, the locations of the domain walls (localization) can be determined.

To realize the HSTE, $\delta$ should be adjusted so that the nonreciprocity is canceled out in the bulk region. We consider parameters that satisfy the fourfold rotation symmetry first; the complex energy spectrum of Fig. 1(a) is shown in Fig. 2(a), and the eigenfield distribution of an edge state is shown in Fig. 2(b).

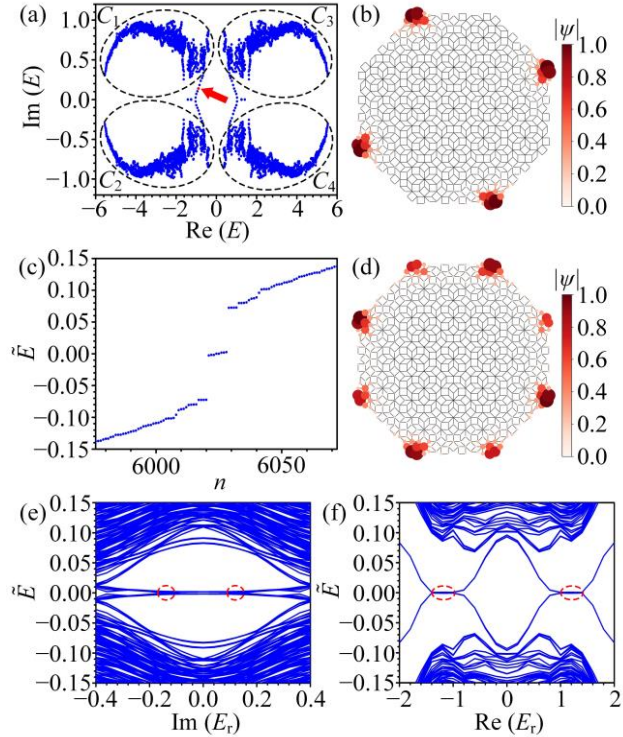

**Fig. 2** (a) The complex energy spectrum for the model in Fig. 1(a), when $\delta_{21} = \delta_{43} = \delta_{65} = \delta_{87} = \delta_{61} = \delta_{25} = \delta_{83} = \delta_{47} = 0.3$, $\delta_{23} = \delta_{45} = \delta_{67} = \delta_{81} = \delta_{27} = \delta_{63} = \delta_{41} = \delta_{85} = 0.8$, and $t = 1.2$ in the Hamiltonian $H$. The Chern numbers for the bulk bands in the dashed circles are $C_1$, $C_2$, $C_3$, and $C_4$. (b) The normalized eigenfield for the eigenvalue $E = -0.8884 + 0.1383i$ [corresponding to the red dot marked by the red arrow in Fig. 2(a)]. (c) The energy spectrum of $\tilde{H}(E_r)$ as a function of the eigenvalue index $n$ for $E_r = -0.8884 + 0.1383i$. (d) The zero-energy eigenfield of $\tilde{H}(E_r)$ in Fig. 2(c). (e) The energy spectrum of $\tilde{H}(E_r)$ as a function of Im $(E_r)$ for Re $(E_r) = -0.8884$. (f) The energy spectrum of $\tilde{H}(E_r)$ as a function of Re $(E_r)$ for Im $(E_r) = 0$. The zero-energy eigenstates are marked by the red dashed ovals.

According to Eq. (4), the Chern numbers for the bulk bands (the assessment of whether the states are bulk states or TESs is provided in Secction D of the Supplementary Material) in Fig. 2(a) are $C_1 = C_4 \approx -1.01$ and $C_2 = C_3 \approx 1.01$. These nontrivial topological invariants ensure the existence of TESs. In Fig. 2(b), it is evident that the eigenfield of the TES is localized at the corners due to the skin effect, i.e., the HSTE is realized.

To further study the skin effect acting on the TESs, the auxiliary Hamiltonian is used for the following analysis. As shown in Fig. 2(c), the energy spectrum of the auxiliary Hamiltonian $\tilde{H}(E_r)$ reveals eight zero-energy eigenstates in the edge gap. Moreover, the eigenfield confirms that these zero-energy eigenstates are corner states, as shown in Fig. 2(d). Specifically, the TES for $E = E_r$ does support the skin effect for the Hamiltonian $H$. In the red dashed circles in Figs. 2(e) and 2(f), the eigenvalues $\tilde{E}$ of $\tilde{H}(E_r)$ in the edge gap are close to 0, indicating the emergence of zero-energy corner states, while the reference energies $E_r$ are consistent with the eigenvalues of the TESs in Fig. 2(a), indicating the realization of the HSTE.

To gain a more comprehensive understanding of the impact of the outer boundary on the HSTE in quasicrystals, we investigate the TESs of quasicrystals for various quadrilateral outer boundaries (the parameters are the same as those in Fig. 2), as shown in Figs. 3(a) and 3(b). The case of a parallelogram outer boundary is discussed in Section E of the Supplementary Material.

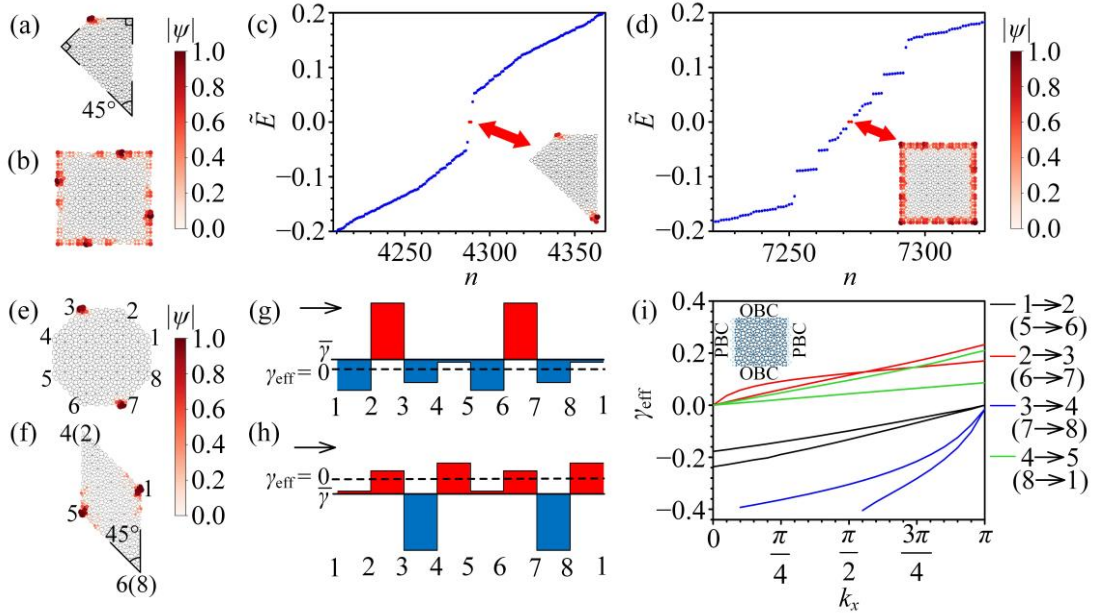

**Fig. 3** (a)–(d) Models with different quadrilateral outer boundaries; $\delta$ and $t$ in the Hamiltonian $H$ are the same as those in Fig. 2(a). Normalized eigenfields of the TESs (the complex energy spectra are shown in Section E of the Supplementary Material): (a) a kite outer boundary and (b) a square outer boundary. The energy spectrum of $\tilde{H}(E_r)$ as a function of the eigenvalue index $n$: (c) $E_r$ is the same as the eigenvalue of Fig. 3(a) for the kite outer boundary, and (d) $E_r$ is the same as the eigenvalue of Fig. 3(b) for the square outer boundary. The zero-energy (marked by the red dots) eigenfields are shown in the insets in the bottom-right corners. (e)–(i) Models with different outer boundaries when $\delta_{21} = \delta_{65} = \delta_{61} = \delta_{25} = 0.4$, $\delta_{23} = \delta_{67} = \delta_{63} = \delta_{27} = 0.8$, $\delta_{43} = \delta_{87} = \delta_{83} = \delta_{47} = 0.6$, $\delta_{45} = \delta_{81} = \delta_{41} = \delta_{85} = 1$, and $t = 0.8$ in the Hamiltonian $H$. Normalized eigenfields of the TESs (the complex energy spectra are shown in Section E of the Supplementary Material): (e) a regular octagon outer boundary (the numbers 1–8 represent the eight corners of the outer boundary, and the correspondence between the corners and the numbers is the same as that below) and (f) a parallelogram outer boundary. [(g), (h)] The two possible inhomogeneous dissipations for the corner localization in Fig. 3(e) (the arrows indicate the directions of chiral currents $v > 0$). (i) Effective dissipation of each open boundary in the $x_{PBC}$–$y_{OBC}$ coordinate system (since the two open boundaries are different, their dissipations are also different, which is represented by two different lines). The definition of $x_{PBC}$–$y_{OBC}$ is illustrated above, and the model for calculating the dissipation is shown in the top-left corner.

In Figs. 2(b) and 3(a), the skin effect acting on the TESs, which manifests itself as a corner localization, can be clearly observed. However, in Fig. 3(b), the distribution of the eigenfield resembles that of the general TES without skin effect (the corner localization of two corners can be achieved for the parallelogram outer boundary; see Section E of the Supplementary Material). To understand whether the skin effect has acted on the TESs in Figs. 3(a) and 3(b), the auxiliary Hamiltonians are studied. As shown in Fig. 3(c), gapped zero-energy eigenstates emerge in the energy spectrum of $\tilde{H}(E_r)$ and, correspondingly, their eigenfields show that the zero-energy eigenstates correspond to corner states. This implies that the skin effect indeed has an impact on the TESs in Fig. 3(a). However, as shown in Fig. 3(d), no gapped zero-energy eigenstates emerge in the energy spectrum of $\tilde{H}(E_r)$, and the eigenfields of the zero-energy eigenstates correspond to edge states. Specifically, the TESs do not permit the occurrence of the skin effect in Fig. 3(b).

To further explore and exploit the localization effects of the HSTE in AB tiling quasicrystals, we change the amplitudes of the nonreciprocal intracell hoppings, which results in the transformation of the fourfold rotation symmetry into a twofold rotation symmetry of a cell, as shown in Fig. 1(b). Unlike Fig. 2(b), Fig. 3(e) only exhibits localization at two corners of the AB tiling quasicrystals for a regular octagon outer boundary due to the variation in the amplitudes of the nonreciprocal intracell hoppings, although the HSTE is also realized. The analysis of the skin effect is presented in Section E of the Supplementary Material. In Fig. 3(f), in contrast with Fig. 3(e), corner localization occurs at corners 1 and 5, indicating that the HSTE established through nonreciprocity also gives rise to domain walls similar to those in the gain–loss system,[54] with these domain walls being associated with the boundary. In Figs. 3(g) and 3(h), the effective dissipation of the edge between two corners [here, the serial number of the two corners are $m$ and $n$; $m, n = 1, \ldots, 8$, corresponding to the serial numbers in Figs. 3(g) and 3(h)] is defined as $\gamma_{\text{eff}(m, n)}$. In Fig. 3(e), the corner localization induced by the inhomogeneous effective dissipation is consistent with Fig. 3(g) [$|\gamma_{\text{eff}(2, 3)}|$ and $|\gamma_{\text{eff}(6, 7)}|$ are significantly larger than the other $|\gamma_{\text{eff}(m, n)}|$] or 3(h) [$|\gamma_{\text{eff}(3, 4)}|$ and $|\gamma_{\text{eff}(7, 8)}|$ are significantly greater than the other $|\gamma_{\text{eff}(m, n)}|$]. The effective dissipation $\gamma_{\text{eff}}$ of each edge is calculated according to Eq. (13) and is shown in Fig. 3(i). The $\gamma_{\text{eff}}$ values in Fig. 3(i) are consistent with those in Fig. 3(h), indicating that the corner localization in Fig. 3(e) is likely to be caused by inhomogeneous dissipation and domain walls similar to those in Fig. 3(h). In Fig. 3(f), the absence of an edge with a large $|\gamma_{\text{eff}}|$ between corners 3 (7) and 4 (8) leads to an increase in $\bar{\gamma}$, resulting in localization at corners 5 and 1, which does not occur in Fig. 3(e). This indicates that by changing the outer boundary, various types of localizations can be realized due to the fact that different outer boundaries exhibit different inhomogeneous dissipations.

By considering the absence of the skin effect in the bulk states of the HSTE model, disorder can be introduced into the nonreciprocal intracell hopping. In a sufficiently large quasicrystal region, the different nonreciprocities caused by disorder can cancel each other out in the bulk region, resulting in the occurrence of the skin effect solely in the TESs. Consequently, the HSTE with disorder can be realized, while the eigenfield may only be localized at one corner occasionally, as shown in Section F of the Supplementary Material. The eight-site cells enable quasicrystals to exhibit various localization phenomena as well as the arbitrary adjustment of the eigenfield distributions of the TESs. This in turn enables localization to take place at different numbers of corners for the same outer boundary [as shown in Figs. 2(b) and 3(e), corner localization occurs at four and two corners, respectively]; it also enables localization to take place at the same number of corners for different outer boundaries [as shown in Figs. 3(e) and 3(f), corner localization occurs at two corners] in quasicrystals. These findings show the potential of the HSTE in tailoring the energy distribution in quasicrystals. See Section G of the Supplementary Material for the case of four-site cells. By contrast, eight-site cells enable a more flexible tuning of the localization.

In summary, we have constructed an HSTE model for AB tiling quasicrystals with eight-site cells and nonreciprocal intracell hoppings, presenting a novel approach for realizing the HSTE in quasicrystals with various polygonal outer boundaries. By changing the outer boundary and adjusting the amplitudes of the nonreciprocal intracell hoppings, we can arbitrarily adjust the localization of TESs in non-Hermitian quasicrystals. Additionally, we discover the existence of domain walls related to the effective dissipation of the outer boundary and present their relationship with the eigenfield distributions of the TESs. These quasicrystals with multi-site cells offer new possibilities for realization of the HSTE and for precise adjustment of the localization.

*Acknowledgement.* This work was supported by the National Natural Science Foundation of China (Grant Nos. 61405058 and 62075059), the Natural Science Foundation of Hunan Province (Grant Nos. 2017JJ2048 and 2020JJ4161), the Scientific Research Foundation of Hunan Provincial Education Department (Grant No. 21A0013), the Open Project of the State Key Laboratory of Advanced Optical Communication Systems and Networks of China (Grant No. 2024GZKF20), and the Guangdong Basic and Applied Basic Research Foundation (Grant No. 2024A1515011353).

# Supplementary Material for "Hybrid Skin-Topological Effect Induced by Eight-Site Cells and Arbitrary Adjustment of the Localization of Topological Edge States"

Jianzhi Chen(陈健智)[1], Aoqian Shi(史奥芊)[1], Yuchen Peng(彭宇宸)[1], Peng Peng(彭鹏)[1], and Jianjun Liu(刘建军)[1,2*]

[1]Key Laboratory for Micro/Nano Optoelectronic Devices of Ministry of Education, School of Physics and Electronics, Hunan University, Changsha 410082, China
[2]Greater Bay Area Institute for Innovation, Hunan University, Guangzhou 511300, China

*Corresponding author. Email: jianjun.liu@hnu.edu.cn

## A. Hamiltonian *H*

Figures 1(c) and 1(d) in the main text represent specific instances of intercell hoppings for a square unit and a rhombic unit, respectively. The signs of the hoppings change with the angles between the cells, aligning with the intracell hoppings (i.e., the two sites with specific numbers connected by a solid or dashed line in a cell, are also connected by a solid or dashed line between cells, respectively), as shown in Fig. A1.

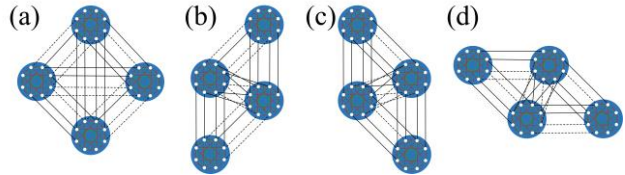

**Fig. A1** (a) The other instances of intercell hoppings for the square unit in Fig. 1(c). The other instances of intercell hoppings for the rhomboid unit in Fig. 1(d): (b)–(d). The solid and dashed lines connecting cells represent positive and negative intercell hoppings, respectively.

According to Eqs. (2) and (3) in the main text, the intracell hopping in a cell in Figs. 1 and A1 is

$$T = \begin{pmatrix} 0 & 1+\delta_{21} & 0 & 1+\delta_{41} & 0 & -1-\delta_{61} & 0 & 1+\delta_{81} \\ 1-\delta_{21} & 0 & 1-\delta_{23} & 0 & 1-\delta_{25} & 0 & 1-\delta_{27} & 0 \\ 0 & 1+\delta_{23} & 0 & -1-\delta_{43} & 0 & 1+\delta_{63} & 0 & -1-\delta_{83} \\ 1-\delta_{41} & 0 & -1+\delta_{43} & 0 & 1-\delta_{45} & 0 & 1-\delta_{47} & 0 \\ 0 & 1+\delta_{25} & 0 & 1+\delta_{45} & 0 & 1+\delta_{65} & 0 & -1-\delta_{85} \\ -1+\delta_{61} & 0 & 1-\delta_{63} & 0 & 1-\delta_{65} & 0 & -1+\delta_{67} & 0 \\ 0 & 1+\delta_{27} & 0 & 1+\delta_{47} & 0 & -1-\delta_{67} & 0 & -1-\delta_{87} \\ 1-\delta_{81} & 0 & -1+\delta_{83} & 0 & -1+\delta_{85} & 0 & -1+\delta_{87} & 0 \end{pmatrix}. \quad (A1)$$

The spatial decay factor of the intercell hopping amplitude is

$$f(r_{jk}) = e^{1-r_{jk}}, \quad (A2)$$

where $r_{jk} = l_0$, $l_1$, or $l_2$. $l_0$ is the side length of the square or rhombus, $l_1$ is the length of the diagonal of the square, and $l_2$ is the length of the short diagonal of the rhombus in Ammann–Beenker (AB) tiling quasicrystals. Here $l_0 = 1$.

The term related to the angle between cells is

$$\Omega(\phi_{jk}) = \begin{pmatrix} 0 & \omega_{\frac{3\pi}{4}}(\phi_{jk}) & 0 & \omega_{\pi}(\phi_{jk}) & 0 & -\omega_{-\frac{3\pi}{4}}(\phi_{jk}) & 0 & \omega_{-\frac{\pi}{2}}(\phi_{jk}) \\ \omega_{-\frac{\pi}{4}}(\phi_{jk}) & 0 & \omega_{\pi}(\phi_{jk}) & 0 & \omega_{-\frac{3\pi}{4}}(\phi_{jk}) & 0 & \omega_{-\frac{\pi}{2}}(\phi_{jk}) & 0 \\ 0 & \omega_0(\phi_{jk}) & 0 & -\omega_{-\frac{3\pi}{4}}(\phi_{jk}) & 0 & \omega_{-\frac{\pi}{2}}(\phi_{jk}) & 0 & -\omega_{-\frac{\pi}{4}}(\phi_{jk}) \\ \omega_0(\phi_{jk}) & 0 & -\omega_{\frac{\pi}{4}}(\phi_{jk}) & 0 & \omega_{-\frac{\pi}{2}}(\phi_{jk}) & 0 & \omega_{-\frac{\pi}{4}}(\phi_{jk}) & 0 \\ 0 & \omega_{\frac{\pi}{4}}(\phi_{jk}) & 0 & \omega_{\frac{\pi}{2}}(\phi_{jk}) & 0 & \omega_{-\frac{\pi}{4}}(\phi_{jk}) & 0 & -\omega_0(\phi_{jk}) \\ -\omega_{\frac{\pi}{4}}(\phi_{jk}) & 0 & \omega_{\frac{\pi}{2}}(\phi_{jk}) & 0 & \omega_{\frac{3\pi}{4}}(\phi_{jk}) & 0 & -\omega_0(\phi_{jk}) & 0 \\ 0 & \omega_{\frac{\pi}{2}}(\phi_{jk}) & 0 & \omega_{\frac{3\pi}{4}}(\phi_{jk}) & 0 & -\omega_{\pi}(\phi_{jk}) & 0 & -\omega_{\frac{\pi}{4}}(\phi_{jk}) \\ \omega_{\frac{\pi}{2}}(\phi_{jk}) & 0 & -\omega_{\frac{3\pi}{4}}(\phi_{jk}) & 0 & -\omega_{\pi}(\phi_{jk}) & 0 & -\omega_{-\frac{3\pi}{4}}(\phi_{jk}) & 0 \end{pmatrix},$$

$$(A3)$$

where

$$\omega_\theta(\phi_{jk}) = \varepsilon_\theta(\phi_{jk})\Phi_\theta(\phi_{jk}), \quad (A4)$$

where $\varepsilon_\theta(\phi_{jk}) = \begin{cases} 1, & |\phi_{jk}-\theta| \leq \frac{\pi}{8} \\ 0, & |\phi_{jk}-\theta| > \frac{\pi}{8} \end{cases}$, $\Phi_\theta(\phi_{jk}) = \cos(\phi_{jk}-\theta)$, $\phi_{jk}$ is the angle between the

hopping direction ($j \rightarrow k$) and the positive horizontal direction.

Moreover, the part of the Hamiltonian related to dissipation $\delta H_D$ in Eq. (12) in the main text is

$$\delta H_D = \sum_j c_j^\dagger \begin{pmatrix} 0 & \delta_{21} & 0 & \delta_{41} & 0 & -\delta_{61} & 0 & \delta_{81} \\ -\delta_{21} & 0 & -\delta_{23} & 0 & -\delta_{25} & 0 & -\delta_{27} & 0 \\ 0 & \delta_{23} & 0 & -\delta_{43} & 0 & \delta_{63} & 0 & -\delta_{83} \\ -\delta_{41} & 0 & \delta_{43} & 0 & -\delta_{45} & 0 & -\delta_{47} & 0 \\ 0 & \delta_{25} & 0 & \delta_{45} & 0 & \delta_{65} & 0 & -\delta_{85} \\ \delta_{61} & 0 & -\delta_{63} & 0 & -\delta_{65} & 0 & \delta_{67} & 0 \\ 0 & \delta_{27} & 0 & \delta_{47} & 0 & -\delta_{67} & 0 & -\delta_{87} \\ -\delta_{81} & 0 & \delta_{83} & 0 & \delta_{85} & 0 & \delta_{87} & 0 \end{pmatrix} c_j. \quad (A5)$$

### B. Kitaev formula

In order to shorten the calculation time, we use another form of the Kitaev formula to calculate the Chern numbers for band $\alpha$:[1]

$$C_\alpha = 12\pi i (\text{Tr}(\bar{A}P^\alpha \bar{B}P^\alpha \bar{C}P^\alpha) - \text{Tr}(\bar{A}P^\alpha \bar{C}P^\alpha \bar{B}P^\alpha)), \quad (B1)$$

where $\bar{A}$, $\bar{B}$ and $\bar{C}$ are the projection operators of regions $A$, $B$ and $C$ in Fig. 1(a), ($\bar{A} + \bar{B} + \bar{C} = \mathbf{1}$, $\mathbf{1}$ is the identity matrix). Under the open boundary condition, it is difficult to completely separate the bulk states from the topological edge states (TESs). In order to avoid the influence of the TESs on the calculation of the Chern numbers for bulk bands, we set the elements of the region close to the outer boundary in the projection operator $P^\alpha$ to be 0, with reference to the calculation of Chern numbers in the hyperbolic model.[2] For the quasicrystal with the regular octagon outer boundary, the distance from the site to the center of the quasicrystal is $d_1$, and the distance from the center of the quasicrystal to its outer boundary is $d_0$. The region close to the outer boundary of the quasicrystal is defined as: $d_1/d_0 > 0.85$.

### C. Local Chern marker

In addition to the Kitaev formula, it is also possible to calculate the real-space Chern numbers in a quasicrystal with the local Chern marker.[3] The local Chern marker is modified below for non-Hermitian systems. In a non-Hermitian system, the local Chern number for band $\alpha$ at the lattice site $r_i$ can be defined as

$$\mathfrak{C}^{\alpha}(r_i) = -4\pi \, \text{Im}\left[\sum_{r_j} \langle r_i | \hat{x}_Q | r_j \rangle \langle r_j | \hat{y}_P | r_i \rangle \right], \quad \text{(C1)}$$

where $\langle r_i | \hat{x}_Q | r_j \rangle = \sum_{r_k} Q^{\alpha}(r_i, r_k) x_k P^{\alpha}(r_k, r_j)$ and $\langle r_j | \hat{y}_P | r_i \rangle = \sum_{r_k} P^{\alpha}(r_j, r_k) y_k Q^{\alpha}(r_k, r_i)$. $P^{\alpha}$ and $Q^{\alpha}$ are the projection operators of a band, where $P^{\alpha}(r_i, r_j) = \sum_{n \in \alpha} \langle r_i | nR \rangle \langle nL | r_j \rangle$ and $Q^{\alpha}(r_i, r_j) = \sum_{n \notin \alpha} \langle r_i | nR \rangle \langle nL | r_j \rangle$. The average Chern number for band $\alpha$ is $C_D^{\alpha} = \frac{1}{\pi r_D^2} \sum_{r_i} \mathfrak{C}^{\alpha}(r_i)$, where $r_D$ is the radius of the selected region $D$.

In order to shorten the calculation time, the local Chern number can be changed to the matrix calculation:

$$C_{\alpha}^{D} = \frac{4\pi i}{\pi r_D^2} \text{Tr}(Q^{\alpha} X P^{\alpha} Y Q^{\alpha}) = \frac{4\pi i}{\pi r_D^2} \text{Tr}(Q^{\alpha} X P^{\alpha} Y Q^{\alpha}), \quad \text{(C2)}$$

where $X$ and $Y$ are coordinate operators for $x$- and $y$-coordinates. Similar to Section B, when tracing, only the elements on the diagonal of the matrix that are not close to the outer boundary are summed (in this case, $d_1/d_0 \leq 0.85$). In Fig. 2(a) in the main text, the average local Chern numbers are $C_1 = C_4 \approx -0.90$ and $C_2 = C_3 \approx 0.90$. It is worth noting that the Chern number studied by the Kitaev formula is more rubust than that studied by the local Chern marker.[3]

### D. Bulk states in quasicrystals

In order to study the Chern numbers for the bulk bands, it is necessary to distinguish the bulk states from the TESs. The method for obtaining pure bulk states is shown in Fig. D1. Since the quasicrystal structure is not periodic, it is impossible to remove all TESs by a periodic boundary condition (PBC) to obtain pure bulk states, but it is still possible to consider the quasicrystal structure as a unit to construct a periodic structure (as shown in Fig. D1(a)) and obtain approximate bulk bands (as shown in Figs. D1(b) and D1(c)).

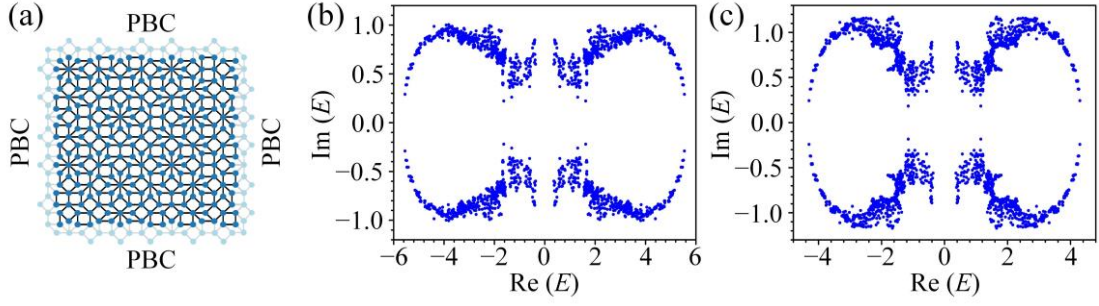

**Fig. D1** (a) Schematic of constructing a periodic structure with the AB tiling quasicrystal (the light-colored region is the equivalent hoppings set for the boundary when constructing a periodic structure). (b) Complex energy spectrum of Fig. D1(a); $\delta$ and $t$ in the Hamiltonian $H$ are the same as those of Fig. 2(a) in the main text. (c) Complex energy spectrum of Fig. D1(a); $\delta$ and $t$ in the Hamiltonian $H$ are the same as those of Fig. 3(e) in the main text.

The eigenstates in the complex energy spectra in Figs. D1(b) and D1(c) correspond to the bulk states without skin effect. According to Figs. D1(b) and D1(c), it is possible to distinguish the bulk states with PBC along both directions from the TESs (see Fig. 2(a) in the main text and Fig. E1 below). In fact, since there is no skin effect for the bulk, it is also possible to distinguish the bulk states from the TESs localized at corners due to skin effect according to the eigenfield distribution.[4–7]

### E. Additional data to Fig. 3 in the main text

For the localization effect of topological states with the quadrilateral outer boundary, the situations for the kite and square outer boundary are analyzed in the main text (as shown in Figs. 3(a) and 3(b)). In order to verify the localization of topological states for the parallelogram outer boundary, the complex energy spectrum of the AB tiling quasicrystal and the eigenfield of its TES for the parallelogram outer boundary are calculated (as shown in Fig. E1(a)). It can be seen that the eigenfield of the TES is localized at the corners due to the skin effect, i.e., the hybrid skin-topological effect (HSTE) is realized.

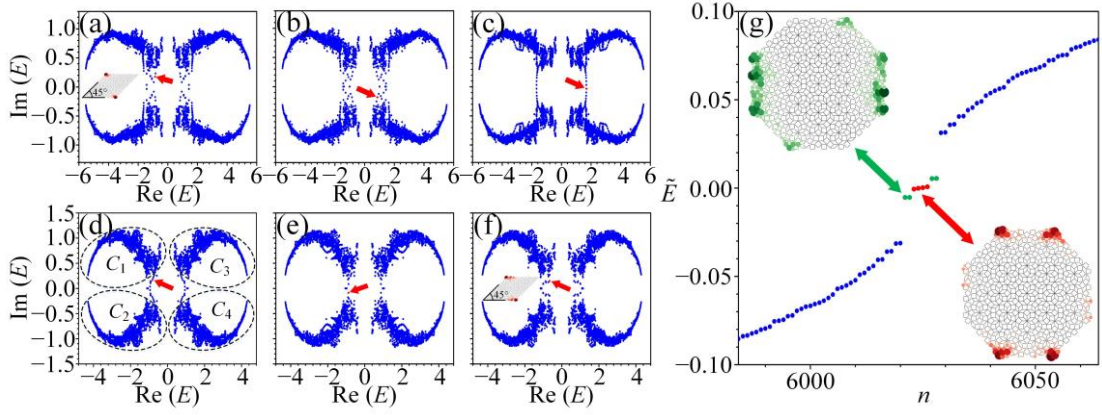

**Fig. E1** (a)–(c) Complex energy spectra of the models with different quadrilateral outer boundaries; $\delta$ and $t$ in the Hamiltonian $H$ are the same as those of Fig. 2(a) in the main text. (a) For the parallelogram outer boundary, the complex energy spectrum and the normalized eigenfield for the red dot marked by the red arrow. (b) and (c) Complex energy spectra of Figs. 3(a) and 3(b) in the main text, where the red dots marked by the red arrows are corresponding to the eigenvalues of the eigenfields in the main text. (d)–(g) Complex energy spectra of the models with different outer boundaries; $\delta$ and $t$ in the Hamiltonian $H$ are the same as those of Fig. 3(e) in the main text. (d) Complex energy spectrum of Fig. 3(e) in the main text. The Chern numbers for the bulk bands in the dashed circles are $C_1$, $C_2$, $C_3$, and $C_4$. The red dot marked by the red arrow is corresponding to the eigenvalue of the eigenfield in the main text. (e) Complex energy spectrum of Fig. 3(f) in the main text, where the red dot marked by the red arrow is corresponding to the eigenvalue of the eigenfield in the main text. (f) Complex energy spectrum and the normalized eigenfield for the red dot marked by the red arrow, when the outer boundary of Fig. 3(f) in the main text is rotated 90°. (g) Energy spectrum of $\tilde{H}(E_r)$ as a function of the eigenvalue index $n$ when $E_r$ is the same as the eigenvalue of Fig. 3(e) in the main text. The zero-energy (marked by the red dots) eigenfield is shown in the inset in the bottom right corner, and the eigenfield for the eigenvalues adjacent to the zero-energy eigenvalues (marked by the green dots) is shown in the inset in the top left corner.

The complex energy spectra of Figs. 3(a), 3(b), 3(e) and 3(f) in the main text are shown in Figs. E1(b)–E1(e), respectively, revealing the emergence of TESs in the bulk gaps. However, the skin effect on TESs cannot be analyzed only according to the complex energy spectrum, and the analysis needs to be combined with the eigenfield or auxiliary Hamiltonian (see the main text). In Fig. E1(d), the Chern numbers for bulk bands are calculated according to Eq. (4) in the main text: $C_1 = C_4 \approx -0.99$ and $C_2 = C_3 \approx 0.99$. The existence of the TESs is ensured by the nontrivial topological invariants. In addition, when the outer boundary of Fig. 3(f) in the main text is rotated 90°, the Hamiltonian for TESs will change because the system is only twofold rotationally symmetric, so the energy spectra of the TESs in Figs. E1(e) and E1(f) are also different.

In the main text, under the action of HSTE, the eigenfield of the TES is localized at two corners in Fig. 3(e). In order to further analyze the HSTE in Fig. 3(e), the auxiliary Hamiltonian is analyzed below. As shown in Fig. E1(g), four degenerate zero-energy eigenstates appear in the edge gap of the auxiliary Hamiltonian $\tilde{H}(E_r)$ (as shown by the red dots). It can be seen from the eigenfield that these zero-energy eigenstates are indeed corner states (as shown in the inset in the bottom right corner in Fig. E1(g)). In other words, the skin effect indeed acts on the TESs of the original Hamiltonian $H$. Different from Fig. 2(c) in the main text, fourfold degenerate zero-energy eigenstates appear in Fig. E1(g), and the corresponding eigenfields are only localized at four corners of the regular octagon, while the eightfold degenerate zero-energy eigenstates appear in Fig. 2(c). Moreover, the four eigenstates adjacent to the zero-energy eigenstates (as shown by the green dots) become edge states (as shown in the inset in the top left corner in Fig. E1(g)). According to these edge states for the auxiliary Hamiltonian, the other four corners are barely affected by the skin effect.

### F. HSTE in disordered cases

For $\omega = 0.2$ or $0.5$ ($\omega$ is the parameter related to disorder), the complex energy spectra of the system with specific $\delta$ and $t$ and the eigenfields for HSTE are shown in Fig. F1.

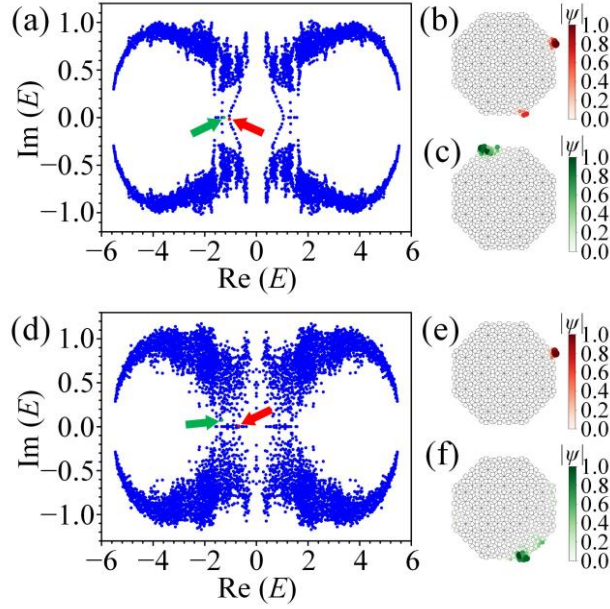

**Fig. F1** Models with the regular octagon outer boundary; $\delta_{21} = \delta_{43} = \delta_{65} = \delta_{87} = \delta_{61} = \delta_{25} = \delta_{83} = \delta_{47} = 0.3 + U$, $\delta_{23} = \delta_{45} = \delta_{67} = \delta_{81} = \delta_{27} = \delta_{63} = \delta_{41} = \delta_{85} = 0.8 + U$, and $t = 1.2$; $U$ is a random number chosen uniformly in the range $[-\omega, \omega]$ in the Hamiltonian $H$. The complex energy spectra; the normalized eigenfields for the red and green dots marked by arrows in the complex energy spectra: (a)–(c) $\omega = 0.2$; (d)–(f) $\omega = 0.5$.

When $\omega = 0.2$ or $0.5$, there are states that realizes HSTE in the bulk gap of the complex energy spectrum (as shown in Figs. F1(a) and F1(d)), and the corner localization of these states may be different (as shown in Figs. F1(b), F1(c), F1(e), and F1(f)). Moreover, considering the above results from the point of view of auxiliary Hamiltonian, when the reference energys $E_r$ are consistent with the eigenvalues of the above TESs, the corner states of the auxiliary Hamiltonian $\tilde{H}(E_r)$ corresponding to Figs. F1(a) and F1(d) are robust to disorder, which is similar to the corner states of the higher-order topological Anderson insulator.[8] It is evident that the HSTE in the quasicrystals can also remain stable for strong disorder, while the corner localization may change for different eigenvalues, which is helpful to further regulate the localization with HSTE.

### G. HSTE with four-site cells in AB tiling quasicrystals

The AB tiling quasicrystals with four-site cells can also realize HSTE, and the analysis is shown in Fig. G1.

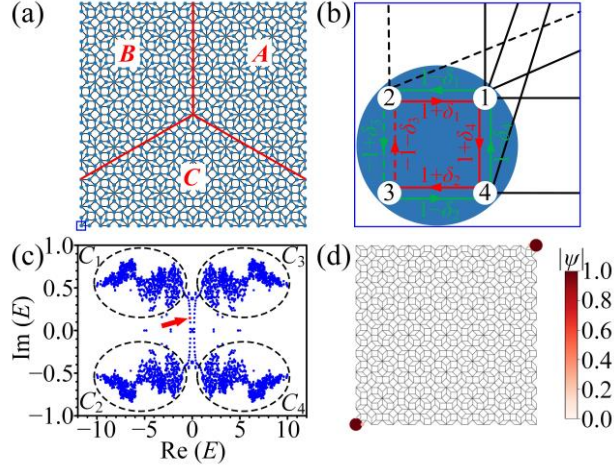

**Fig. G1** (a) AB tiling quasicrystal with a square outer boundary. The inner boundary is set for calculating the Chern numbers for the bulk bands with the Kitaev formula. The three regions divided by the inner boundary are marked by $A$, $B$, and $C$. (b) Schematic of the four-site cell and the hoppings (here is a magnification of the blue box in the bottom left corner in Fig. G1(a)). When $\delta_1 = \delta_2 = 0.3$, $\delta_3 = \delta_4 = -1$, $t = 0.85$: (c) the complex energy spectrum. The Chern numbers for the bulk bands in the dashed circles are $C_1$, $C_2$, $C_3$, and $C_4$; (d) the normalized eigenfield for the red dot marked by the red arrow in Fig. G1(c).

As shown in Fig. G1(a), a square outer boundary is set for the quasicrystal, where each point represents a cell and each cell contains four sites (see Fig. G1(b)). The intercell hoppings are introduced on the sides of the square units, as well as the sides and shorter diagonals of the rhomboid units (similar to Figs. 1(c) and 1(d) in the main text, but no hoppings are introduced on the diagonals of the square units, different from the main text). Analog to the two-dimensional square lattice Su–Schrieffer–Heeger (SSH) model realizing HSTE,[4] the nonreciprocal intracell hoppings are introduced (the nonreciprocity is achieved using a non-zero $\delta$, with $\delta$ quantifying the nonreciprocal hopping between two sites).

The form of Hamiltonian $H$ of the system is the same as that in the main text (i.e., Eqs. (1), (2), and (3)), but the specific contents of $T$, $t$, $f(r_{jk})$, and $\Omega(\phi_{jk})$ are different:

$T$ is the intracell hopping of a cell,

$$T = \Gamma_1 + \Gamma_2 + i\frac{\delta_1+\delta_2}{2}\Gamma_3 + i\frac{\delta_1-\delta_2}{2}\Gamma_4 - i\frac{\delta_3+\delta_4}{2}\Gamma_5 + i\frac{\delta_3-\delta_4}{2}\Gamma_6, \tag{G1}$$

where $\Gamma_1 = \tau_2\tau_2$, $\Gamma_2 = \tau_0\tau_1$, $\Gamma_3 = \tau_0\tau_2$, $\Gamma_4 = \tau_3\tau_2$, $\Gamma_5 = \tau_2\tau_1$, and $\Gamma_6 = \tau_1\tau_2$. $\tau_{1,2,3}$ are the Pauli matrices, and $\tau_0$ is the identity matrix.

$t$ is the intercell hopping amplitude, and $f(r_{jk})$ is the spatial decay factor of the intercell hopping amplitude,

$$f(r_{jk}) = e^{1-r_{jk}}, \tag{G2}$$

where $r_{jk} = l_0$ or $l_1$. $l_0$ is the side length of the square or rhombus, and $l_1$ is the length of the short diagonal of the rhombus in the AB tiling quasicrystals. Here $l_0 = 1$.

$\Omega(\phi_{jk})$ is a term related to the angle between cells,

$$\Omega(\phi_{jk}) = \cos\phi_{jk}\Gamma_2 - |\cos\phi_{jk}|i\Gamma_4 - \sin\phi_{jk}\Gamma_1 - |\sin\phi_{jk}|i\Gamma_6, \tag{G3}$$

where $\phi_{jk}$ is the angle between the hopping direction ($j\rightarrow k$) and the positive horizontal direction.

According to Eq. (4) in the main text, the Chern numbers for the bulk bands in Fig. G1(c) are $C_1 = C_4 \approx -0.99$ and $C_2 = C_3 \approx 0.99$, and the nontrivial topological invariants ensure the existence of the TESs. In Fig. G1(d), the eigenfield of the TES is localized at the corners due to the skin effect, i.e., the HSTE is realized.

245302